\documentstyle[12pt,fbsty,epsf,epsfig]{article}

\epsfverbosetrue

 1
 1
 1

\def\bbuildrel#1_#2^#3{\mathrel{\mathop{\kern 0pt#1}\limits_{#2}^{#3}}}
\def\slash#1{\setbox0=\hbox{$#1$}#1\hskip-\wd0\dimen0=5pt\advance
       \dimen0 by-\ht0\advance\dimen0 by\dp0\lower0.5\dimen0\hbox
         to\wd0{\hss\sl/\/\hss}}
\def\beq{\begin{equation}}
\def\eeq{\end{equation}}

\newcommand{\be}{\begin{equation}}
\newcommand{\ee}{\end{equation}}
\newcommand{\f}{\frac}

\newcommand{\e}{\epsilon}

\newcommand{\smallw}{{\scriptscriptstyle W}}

\newcommand{\mw}{M_\smallw}
\newcommand{\muw}{\mu_\smallw}

\newcommand{\as}{\alpha_{\scriptscriptstyle S}}

\newcommand{\smallsm}{\rm{SM}}

\newcommand{\smallthdm}{\rm{H}}

\newcommand {\ms } {$\overline{\rm{MS}}$ } 
\begin{document}
\begin{titlepage}

 \begin{flushright}
  {
      hep-ph/0407323\\
}
 \end{flushright}

 \begin{center}

\setlength {\baselineskip}{0.3in}
{\large
{\bf QCD corrections to the Wilson coefficients {\boldmath $C_9$} and 
{\boldmath $C_{10}$}
in two-Higgs-doublet models}}
\vspace{1.5cm} \\
\setlength {\baselineskip}{0.2in}

{\large  S. Schilling}\\
\vspace{0.5cm}
{\it Institut f\"ur Theoretische Physik, 
 Universit\"at Z\"urich,                          \\
 Winterthurerstrasse 190, 
 8057 Z\"urich, Switzerland}                     \\[1.01ex]

\vspace{0.7cm}
{\large  C. Greub, N. Salzmann and B. T\"odtli}\\
\vspace{0.5cm}
{\it Institut f\"ur Theoretische Physik, 
 Universit\"at Bern,                       \\
  Sidlerstrasse 5, 3012 Bern,
  Switzerland}  
\vspace{1.7cm}

{\bf Abstract \\}
\end{center}
In this letter we present the analytic results for the two-loop 
corrections to the Wilson coefficients $C_{9}(\mu_W)$ and $C_{10}(\mu_W)$ 
in type-I and
type-II two-Higgs-doublet models at the matching scale $\mu_W$. 
These corrections are important ingredients for 
next-to-next-to-leading logarithmic predictions of various observables
related to the decays $B \to X_s l^+ l^-$ in these models. 
In scenarios with moderate values of $\tan \beta$ neutral
Higgs boson contributions can be safely neglected for $l=e,\mu$.
Therefore we concentrate on the contributions mediated by charged
Higgs bosons.

\setlength{\baselineskip}{0.3in} 

\end{titlepage} 

\setlength{\baselineskip}{0.3in}

\section{Introduction}
\label{intro}
In the standard model (SM) rare decays of $B$-mesons like $B \to X_{s,d} 
\gamma$ or $B \to X_{s,d} l^+ l^-$ are induced by one-loop diagrams. 
In many extensions
of the SM, there are additional one-loop contributions
in which non-SM particles propagate in the loop. If the new particles
are not considerably heavier than those of the SM,
the new contributions to these decays can be as large as
the SM ones. As an illustration of the high sensitivity of these decays
to new physics, we mention that the most stringent bound on the
mass of the charged Higgs-boson in the type-II two-Higgs-doublet model 
comes from rare $B-$decays, viz $B \to X_s \gamma$, leading to 
$M_H > 280$ GeV ($99\%$ C.L.) \cite{Gambino:2001ew}.

It goes without saying that one should try to get information on the
parameters in a given extension -- 
the two-Higgs-doublet models in this letter -- 
from all processes which allow  both a clean theoretical prediction and an 
accurate measurement. This means that 
precision studies similar to those for $B \to X_s \gamma$ 
\cite{Gambino2,BorzumatiGreub,Strumia,MU00},
where higher order QCD corrections are crucial,
should also be done for the process $B \to X_s l^+ l^-$. On the theoretical
side this means that next-to-next-to leading logarithmic (NNLL) calculations
for the branching ratio and/or the forward-backward asymmetry
are needed.

In this letter we consider QCD corrections to the process $B \to X_s l^+ \l^-$
($l=e,\mu$) in 2HDMs. We neglect diagrams with neutral Higgs-boson exchange.
This omission is justified
in the type-II model, if the coupling parameters
$(m_l/M_W) \tan \beta$ and
$m_l/(M_W \cos \beta)$ are sufficiently smaller than one. 
In this case the operator basis is the same as in the SM.
Only the matching calculation for the
Wilson coefficients gets changed by adding the contributions where the
flavor transition is mediated by the exchange of the physical charged Higgs
boson. While these extra pieces are known for the
coefficients $C_7$ and $C_8$ to two-loop precision 
for quite some time,
the corresponding results for 
$C_9$ and $C_{10}$, presented in this letter, were not published before.
The phenomenological consequences for the branching
ratio and other observables will be discussed in \cite{Schilling_long}.

The remainder of this letter is organized as follows:
In section 2, we summarize the necessary  aspects of the 2HDMs.
In section 3 we first present the effective Hamiltonian, followed by the
analytic results 
for the charged Higgs boson contributions
to the Wilson coefficients $C_9(\mu_W)$ and $C_{10}(\mu_W)$. 
In this section we also briefly 
investigate how the two-loop corrections reduce the renormalization scheme
dependence related to the definition of the top-quark mass.

\section{Two-Higgs-doublet models}
\label{2hdm}
In the following we consider models with two complex Higgs-doublets 
$\phi_1 $ and $ \phi_2$.
After spontaneous symmetry breaking these two doublets give rise to 
two charged ($H^{\pm}$) and three neutral ($H^0$, $h^0$, $A^0$) 
Higgs-bosons.
When requiring the
absence of flavour changing neutral currents at the tree-level, as we
do in this paper, one obtains two possibilites, the type-I and the type-II
2HDM \cite{Weinberg}. 
The part of the Lagrangian relevant for our calculation is
the Yukawa interaction between the charged physical Higgs bosons 
$H^{\pm}$ and the quarks (in its mass eigenstate basis):
\beq 
{\cal L_I} =\frac{\,g}{\sqrt{2}} \left\{
\left(\frac{{m_d}_i}{\mw}\right)
      X \,{\overline{u}_L}_j V_{ji}  \, {d_R}_i+
\left(\frac{{m_u}_i}{\mw}\right)
      Y \,{\overline{u}_R}_i V_{ij}  \, {d_L}_j
                               \right\} H^+
 +{\rm h.c.}\,.
\label{higgslag}
\eeq
The couplings $X$ and $Y$ are 
\bea
\begin{array}{llll}
X=-\cot \beta , & Y &=&\cot \beta \qquad \mbox{(type-I)}, \nonumber \\
X=  \;\;\tan \beta ,& Y &=&\cot \beta \qquad \mbox{(type-II)},
\end{array}
\eea
where $\tan \beta= v_2/v_1$, with $v_1$ and $v_2$ being the vacuum
expectation values of the Higgs doublets $\phi_1$ and $\phi_2$, respectively.

In the following we will use the generic form (\ref{higgslag})
for the interaction between
$H^{\pm}$ and the quarks. It will turn out that the Wilson 
coefficients $C_9(\mu_W)$ and $C_{10}(\mu_W)$ are independent of the 
model (type-I or type-II), as they only depend on $Y^2$.

\section{Charged Higgs contributions to {\boldmath $C_9(\mu_W)$} and 
{\boldmath $C_{10}(\mu_W)$} at the two-loop level}
\label{matching}
In this section we first briefly describe the effective Hamiltonian.
We then present the analytic results up to two loops for the charged Higgs boson
contributions to $C_9(\mu_W)$ and $C_{10}(\mu_W)$. Finally we briefly
investigate the impact of the new two-loop contributions on $C_9(\mu_W)$. 
\subsection{Effective Hamiltonian}
To describe decays like $B \to X_s l^+ l^-$ we use the framework 
of an effective low--energy theory with five
quarks, obtained by integrating out the heavy degrees of freedom.
In the present case these are the $t$-quark, the $W^{\pm}$ and $Z^0$
boson as well as the charged Higgs bosons $H^{\pm}$, whose masses $M_H$ are
assumed to be of the same order of magnitude as $M_W$.
As in the SM calculations we only take
into account operators up to dimension six and set $m_s=0$.
In  these approximations the   
effective Hamiltonian relevant for our application
(with $ |\Delta B| = |\Delta S| =1$)  
\be
 {\cal H}_{eff} = - \frac{4 G_F}{\sqrt{2}} \,V_{ts}^\star V_{tb} 
   \sum_{i=1}^{10} C_i(\mu) {\cal O}_i(\mu)  
\ee
contains precisely the same operators ${\cal O}_i(\mu)$
as in the SM case. They read:
\beq
\begin{array}{llll}
{\cal O}_1 \,= &\!
 (\bar{s}_L \gamma_\mu T^a c_L)\, 
 (\bar{c}_L \gamma^\mu T^a b_L)\,, 
               &  \quad 
{\cal O}_2 \,= &\!
 (\bar{s}_L \gamma_\mu c_L)\, 
 (\bar{c}_L \gamma^\mu b_L)\,,   \\[1.002ex]
{\cal O}_3 \,= &\!
 (\bar{s}_L \gamma_\mu b_L) 
 \sum_q
 (\bar{q} \gamma^\mu q)\,, 
               &  \quad 
{\cal O}_4 \,= &\!
 (\bar{s}_L \gamma_\mu T^a b_L) 
 \sum_q
 (\bar{q} \gamma^\mu T^a q)\,,  \\[1.002ex]
{\cal O}_5 \,= &\!
 (\bar{s}_L \gamma_\mu \gamma_\nu \gamma_\rho b_L) 
 \sum_q
 (\bar{q} \gamma^\mu \gamma^\nu \gamma^\rho q)\,, 
               &  \quad 
{\cal O}_6 \,= &\!
 (\bar{s}_L \gamma_\mu \gamma_\nu \gamma_\rho T^a b_L) 
 \sum_q
 (\bar{q} \gamma^\mu \gamma^\nu \gamma^\rho T^a q)\,,  \\[1.002ex]
{\cal O}_7 \, = & \!
 \f{e}{g_s^2} m_b (\bar{s}_L \sigma^{\mu \nu}     b_R) F_{\mu \nu}\,, &\quad
 {\cal O}_8 \, = & \! \f{1}{g_s} m_b (\bar{s}_L \sigma^{\mu \nu} T^a b_R) 
G_{\mu \nu}^a\, ,
 \\[1,00ex]
{\cal O}_9   \,= &\! \f{e^2}{g_s^2} (\bar{s}_L \gamma_{\mu} b_L) \sum_l 
                                      (\bar{l}\gamma^{\mu} l)\,,
  &  \quad 
{\cal O}_{10}  \,= &\! \f{e^2}{g_s^2} (\bar{s}_L \gamma_{\mu} b_L) \sum_l 
                             (\bar{l} \gamma^{\mu} \gamma_5 l),  

\end{array} 
\label{opbasis}
\eeq
where $T^a$ ($a=1,...,8$) are the $SU(3)$ colour generators, and
$g_s$ and $e$ are the strong and electromagnetic coupling constants. 
$q$ and $l$ appearing in the sums run over the light quarks ($q=u,...,b$) and
the charged leptons, respectively.

 The Wilson coefficients $C_i(\mu)$ are found in the matching procedure by
requiring that conveniently chosen Green's  functions or on-shell
matrix elements are equal when
calculated in the effective theory and in the underlying full theory up
to ${\cal O}[($external momenta and light masses$)^2/M^2]$, where $M$ 
denotes one of the heavy masses like $M_W$ or $ M_H$.
The matching scale $\mu_W$ is usually chosen to be at the order of 
$M$, because at this scale
the matrix elements or Green's functions of the effective operators  pick up the  same large
logarithms as the corresponding quantities in the full theory. 
Consequently, the Wilson coefficients 
$C_i(\muw)$ only pick up ``small'' QCD corrections,
which can be calculated in fixed order perturbation theory.
For the following it is convenient to expand the Wilson $C_i(\mu_W)$ as
\be 
\label{expanded.coeffs}
C_i(\mu_W) = C^{(0)}_i(\mu_W) 
+ \f{g_s^2}{(4 \pi)^2} C^{(1)}_i(\mu_W)  
+ \f{g_s^4}{(4 \pi)^4} C^{(2)}_i (\mu_W) 
+ {\cal O}(g_s^6).
\ee
We note that due to the particular convention concerning the
powers of the strong
coupling constant $g_s$ in the definition of our operators, the 
contributions of order $g_s^{2n}$ to each Wilson coefficient originate
from $n$-loop diagrams. 

In the SM all the Wilson coefficients $C_i(\mu_W)$ are known
at the two-loop level.
In 2HDMs, the charged Higgs boson exchanges lead to additional
contributions. For the following discussion,
we split the Wilson coefficients into a SM- and charged
Higgs boson contribution  according to
\be
\label{expanded.higgscoeffs}
C_{\,_i}(\muw)=C_{i,\smallsm}(\mu_W) +  C_{i,\smallthdm}(\mu_W) \, .
\ee
The individual pieces $C_{i,\smallsm}(\mu_W)$ and
$C_{i,\smallthdm}(\mu_W)$ can be expanded in $g_s$
in the same way as $C_i(\mu_W)$ in eq. (\ref{expanded.coeffs}). 
While $C_{7,\smallthdm}(\mu_W)$ and  $C_{8,\smallthdm}(\mu_W)$ are known 
at the
two-loop level \cite{Gambino2,BorzumatiGreub,MU00}, 
$C_{9,\smallthdm}(\mu_W)$ and $C_{10,\smallthdm}(\mu_W)$ were up to 
now only known to one-loop precision \cite{Masiero,WylerCho}. 

\subsection{Analytic results for {\boldmath $C_{9,\smallthdm}^{(2)}(\mu_W)$}
  and {\boldmath $C_{10,\smallthdm}^{(2)}(\mu_W)$}}
\label{results}
\begin{figure}[htb]
\begin{center}
\epsfig{file=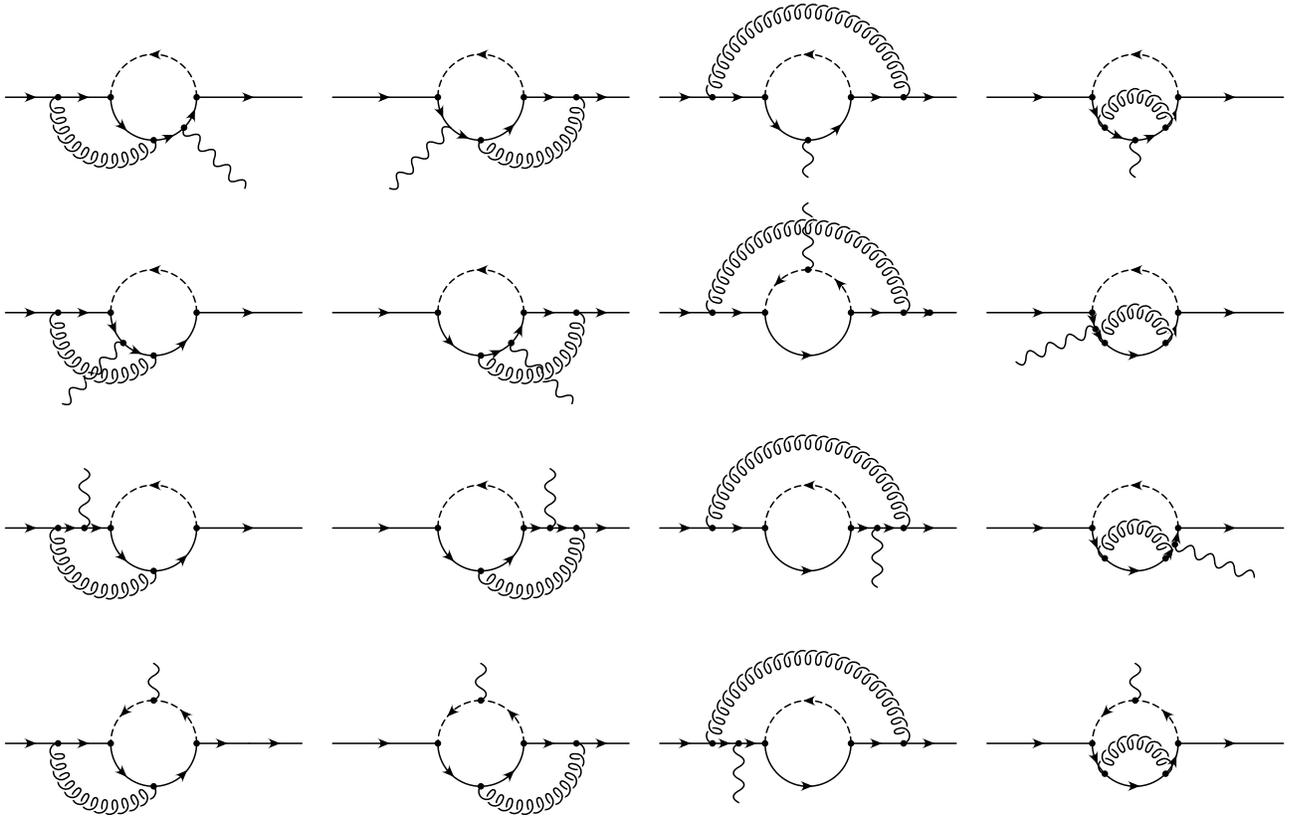,width=17cm}
\end{center}
\caption{One-particle irreducible two-loop diagrams for 
$b \to s l^+ l^-$ needed to extract the charged Higgs boson contribution to
$C_{9}^{(2)}(\mu_W)$ and $C_{10}^{(2)}(\mu_W)$. 
The external quark lines (solid) denote the incoming $b$-quark and the
outgoing $s$-quark, while the wavy line denotes a virtual photon or a
$Z^0$-boson, which decays into a $l^+l^-$-pair. The internal dashed-, solid-
and curly lines denote the charged Higgs boson $H^\pm$, the $t$-quark and the
gluon, respectively.}
\label{feyndiag}
\end{figure}

We did the matching calculation for $C_{9,\smallthdm}^{(2)}(\mu_W)$ 
and $C_{10,\smallthdm}^{(2)}(\mu_W)$ in two different ways, leading to identical final
results: On the one hand we performed a matching calculation for
(the off-shell) Green's function related to $b \to s l^+ l^-$, as described in
detail for the SM in \cite{MU99}. 
On the other hand we matched the corresponding on-shell
 amplitude onto the effective theory, following
basically the methods described in \cite{GreubHurth},
 but using some simplifications\footnote{As a byproduct of our calculation, 
we also confirmed
the known result for the charged Higgs
contribution  $C_{7,\smallthdm}^{(2)}$ 
(see e.g. \cite{Gambino2,BorzumatiGreub}).}.
In both methods, the hard part of the calculation consists of working 
out the one-particle irreducible diagrams shown in fig. \ref{feyndiag}. 

After using heavy mass expansion techniques \cite{Smirnov}, partial fraction 
decomposition and the usual reduction of tensor integrals to scalar ones, 
we obtain integrals of the type 
\bea
C^{(2)}_{n_1 n_2 n_3} &=& 
\f{(m_1^2)^{n_1+n_2+n_3-4+2\e}}{\pi^{4-2\e}\; \Gamma(1+\e)^2}
\int \f{d^{4-2\e}q_1 \; d^{4-2\e}q_2}{(q_1^2 - m_1^2)^{n_1}
                  (q_2^2 - m_2^2)^{n_2}[(q_1 - q_2)^2]^{n_3}},
\label{int2}
\eea
which are known explicitly \cite{DT93,vBG}.

Note that only the contributions from the internal
top-quarks have to be taken into account in these diagrams, because the charm
contributions, which
come with a relative suppression factor of 
$m_b m_c/m_t^2$ or $m_c^2/m_t^2$, only induce dimension 8 operators 
which are neglected in our treatment.

 We write the one- and two-loop charged Higgs induced
contributions to $C_9(\mu_W)$ and $C_{10}(\mu_W)$ in the form
\bea
C_{9,\smallthdm}^{(n)}(\muw)&=&Y^2 \,\left(\Gamma^{(n)} + \;
\f{1-4s_W^2}{s_W^2 }\, 
\,Z^{(n)}\right) \, , \nonumber \\
C_{10,\smallthdm}^{(n)}(\muw)&=&- Y^2 \, \left(\;\f{1}{s_W^2} \,
  Z^{(n)}\right) \, ,
\label{de}
\eea
where $s_W=\sin \theta_W$.
The terms proportional to $Z^{(n)}$ ($\Gamma^{(n)}$) account for
the $n$-loop $Z^0$- (photon-) penguin diagrams.

\noindent
The one-loop contributions 
$\Gamma^{(1)}$ and  $Z^{(1)}$ 
read \cite{WylerCho}
\bea
\Gamma^{(1)}
&=&\frac{-\left( 38 - 79\,y + 47\,y^2 \right) y
           }{108\,{\left(  y -1 \right) }^3} + 
  \frac{\left( 4 - 6\,y + 3\,y^3 \right) y}
   {18\,{\left( y -1 \right) }^4}\ln y ,\nonumber \\
Z^{(1)}&=&\frac{xy}{8\,\left( y -1 \right) } - 
  \frac{xy}{8\,{\left(  y -1 \right) }^2}\ln y,
\label{oneloop}
\eea
with 
\be
\label{xy}
x=\frac{m_t^2}{M_W^2} \quad ; \quad y =\frac{m_t^2}{M_H^2} \, .
\ee
Note that $\Gamma^{(1)}$ and $Z^{(1)}$ depend via $x$ and $y$ on the
renormalization scheme for the $t$-quark mass.
To illustrate this dependence, we give our results in the commonly used 
$\overline{\rm{MS}}$- and pole-mass scheme. 
The relation between these mass definitions is given by
\beq
 {\overline m}_t(\mu_W) = m_t^{\rm{pole}}
 \left(1 + \frac{2 \as(\mu_W)}{\pi} \, \ln \frac{m_t^{\rm{pole}}}{\mu_W} 
- \frac{4}{3} \frac{\as(\mu_W)}{\pi}\right) + {\cal O}(\alpha_s^2) \, ,
\label{polerunning}
\eeq
where $\overline{m}_t(\mu_W)$ and $m_t^{\rm{pole}}$ are the top-quark mass 
in the $\overline{\rm{MS}}$-scheme and pole-mass scheme, respectively. 

The new two-loop terms 
$\Gamma^{(2)}$  and $Z^{(2)}$, which explicitly depend on the top-mass 
renormalization scheme, can be written as
\begin{eqnarray}
\label{oldms}
\Gamma^{(2)}&=& 
W_{\Gamma} + N_{\Gamma} \ln \frac{{\mu_W}^2}{M_H^2}
+ \left(\ln \f{m_t^2}{\mu_W^2}-\f{4}{3}\right) T_\Gamma  \, ,
\nonumber \\
Z^{(2)} &=&
W_{Z}+N_{Z} \ln \frac{{\mu_W}^2}{M_H^2}
+ \left(\ln \f{m_t^2}{\mu_W^2}-\f{4}{3}\right) T_Z \, .
\end{eqnarray}
The expressions for $W_\Gamma$, $N_\Gamma$, $W_Z$ and $N_Z$ are the
same in both schemes (up to the different $m_t$ in the definition of $x$ and
$y$):
\bea
W_{\Gamma}&=&\frac{\left( 764 + 3927\,y - 9138\,y^2 + 6175\,y^3 \right) y}
{729\,{\left(  y -1 \right) }^4} \nonumber \\ &&
- \frac{4\left( 32 + 18\,y - 132\,y^2 + 95\,y^3 \right) y}{81\,{\left(  
y -1 \right) }^4}  {\rm Li}_2\left(\frac{y-1}{y}\right) 
\nonumber\\ 
&& 
- \frac{4\,\left( -110 + 797\,y - 1233\,y^2 + 602\,y^3 + 88\,y^4 \right) y}
{243\,{\left(  y -1 \right) }^5} \,\ln y 
\nonumber\\ 
&& 
+\frac{8\,\left( 16 + 5\,y - 57\,y^2 + 54\,y^3 \right) y}
{81\,{\left(  y -1 \right) }^5} \, \ln^2 y ,
\label{wgamma} \nonumber \\
N_{\Gamma}&=&\frac{4\,\left( 263 - 486\,y + 243\,y^2 + 88\,y^3 \right) y}{243\,
{\left(  y -1 \right) }^4} - 
\frac{8\,\left( 16 + 5\,y - 57\,y^2 + 54\,y^3 \right) y}{81\,{
\left(  y -1 \right) }^5}\ln y,
\label{ngamma} \nonumber \\
W_Z&=&\frac{4\left( -3 +y \right) x y}{3{\left( y -1 \right) }^2}
   + \f{\left( 2 -y\right) x y }
   {{\left( y  -1 \right) }^2} {\rm Li}_2 \left(\f{y-1}{y}\right) \nonumber \\ &&   
+ \frac{\left( 2 + 9\,y - 3\,y^2 \right) x y }{3{\left(y -1 \right) }^3} \ln y - 
\f{2xy}{{\left( y -1\right) }^3} \ln^2 y ,
\label{wz} \nonumber \\
N_Z&=& \f{( -3 + y) xy }
  {(  y  -1) ^2} + 
     \f{2 xy} {(y -1)^3}\ln y \, ,
\label{nz}
\eea
where the function ${\rm Li}_2(z)$  is defined as 
\bea
{\rm Li}_2(z) &=& -\int_0^z \f{dt}{t} \ln (1-t) \, .
\eea
The expressions for $T_\Gamma$ and $T_Z$ depend
on the renormalization scheme used for $m_t$:

\bea
\overline{\rm{MS}}-\rm{scheme}:  T_\Gamma&=&0,\nonumber \\
                    T_Z    &=&0,\nonumber\\
\rm{pole -scheme}:T_\Gamma&=&
\frac{4\,\left( 31 - 59\,y + 31\,y^2 + 9\,y^3 \right) y}
   {27\,{\left( y  -1\right) }^4}
-\frac{16\,\left( 1 - 3\,y^2 + 3\,y^3 \right) y}
   {9\,{\left( y  -1\right) }^5}\ln y, \, \nonumber \\  
T_Z&=& \frac{(3 - 4\,y + y^2)x y}{{\left(  y -1 \right) }^3} +
  \frac{2xy}{{\left(  y -1 \right) }^3}\ln y.
\eea

\subsection{Impact of the two-loop contributions on {\boldmath $C_{10,\smallthdm}$}}
\begin{figure}[htb]
\begin{center}
\epsfig{file=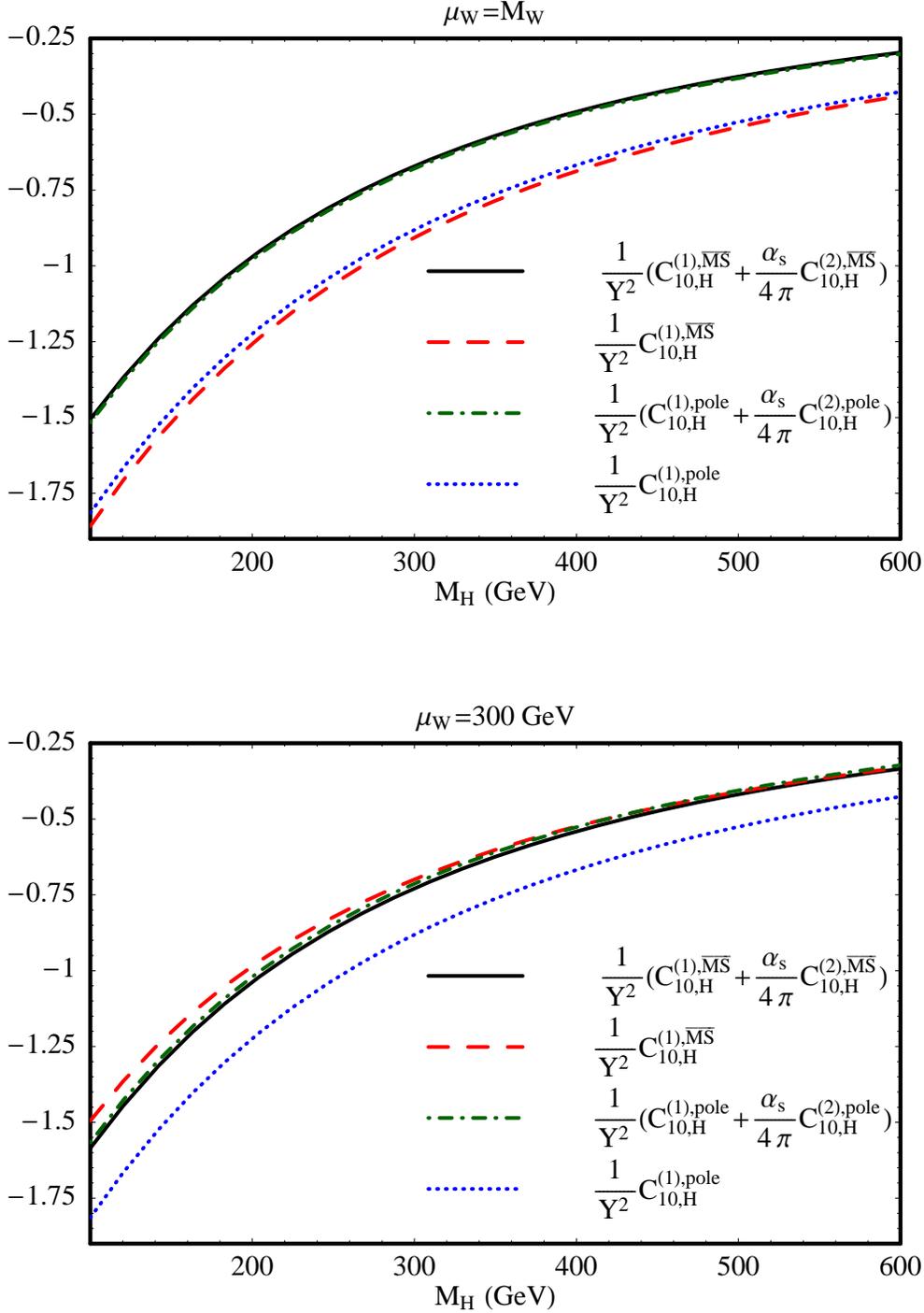,width=15.5cm}
\end{center}
\caption{Dependence of the rescaled Wilson coefficient 
$\hat{C}_{10,\smallthdm}(\mu_W)$ 
(see eq. (\ref{rescaled})) on the charged Higgs boson mass $M_H$
at the matching scale
$\mu_W=M_W$ (upper frame) and $\mu_W=300$ GeV (lower frame).
The dashed (dotted) line is the one-loop contribution expressed in
\ms -scheme (pole-mass scheme) of the $t$-quark mass, 
while the solid (dash-dotted) line includes the two-loop corrections  in the
respective scheme.}
\label{combi}
\end{figure}

In this section we briefly illustrate the impact of the two-loop corrections presented
in this letter on $C_{10,\smallthdm}(\mu)$. 
We introduce a rescaled Wilson coefficient (see eq. (\ref{expanded.coeffs}))
\begin{equation}
\label{rescaled}
\hat{C}_{10,\smallthdm}(\mu_W) \doteq \frac{1}{Y^2} \,
\frac{4\pi}{\alpha_s(\mu_W)} 
C_{10,\smallthdm}(\mu_W),
\end{equation}
In fig. \ref{combi}
we plot the quantities
\be
\frac{1}{Y^2} C_{10,\smallthdm}^{(1)}(\mu_W) \quad \mbox{and} \quad
\frac{1}{Y^2} \left( C_{10,\smallthdm}^{(1)}(\mu_W) + 
\frac{\alpha_s(\mu_W)}{4\pi}
                     C_{10,\smallthdm}^{(2)}(\mu_W) \right)  ,
\ee
i.e. two approximations of $\hat{C}_{10,\smallthdm}$
as a function of the charged Higgs boson mass $M_H$
for the $\overline{\rm{MS}}$- and for the pole mass scheme of the $t$-quark
mass. As input parameters we use 
$\alpha_s(M_Z) = 0.119$,
$m_t^{\rm{pole}} = 178.0~{\rm GeV}$,   
$M_W = 80.4~{\rm GeV}$ and         
$s^2_W = 0.231$ \cite{PData02,nature}.
The upper frame shows these quantities at the relatively low matching scale
$\mu_W=M_W$. As in this case $m_t^{\rm{pole}}$ and $\overline{m}_t(\mu_W)$ 
are numerically
almost identical, the one-loop approximations (dotted and dashed lines) 
are close
to each other. The inclusion of the two-loop corrections, however, 
considerably lowers
the (absolute) size of the coefficient for all values of $M_H$ considered. 
In the lower
frame a higher matching scale of $\mu_W=300$ GeV is chosen. As in this case
$m_t^{\rm{pole}}$ and $\overline{m}_t(\mu_W)$ differ considerably, 
the renormalization
scheme dependence of the one-loop results is rather large. When taking into
account the two-loop corrections (solid and dash-dotted lines), 
the scheme dependence is drastically reduced.
\begin{figure}[htb]
\begin{center}
\epsfig{file=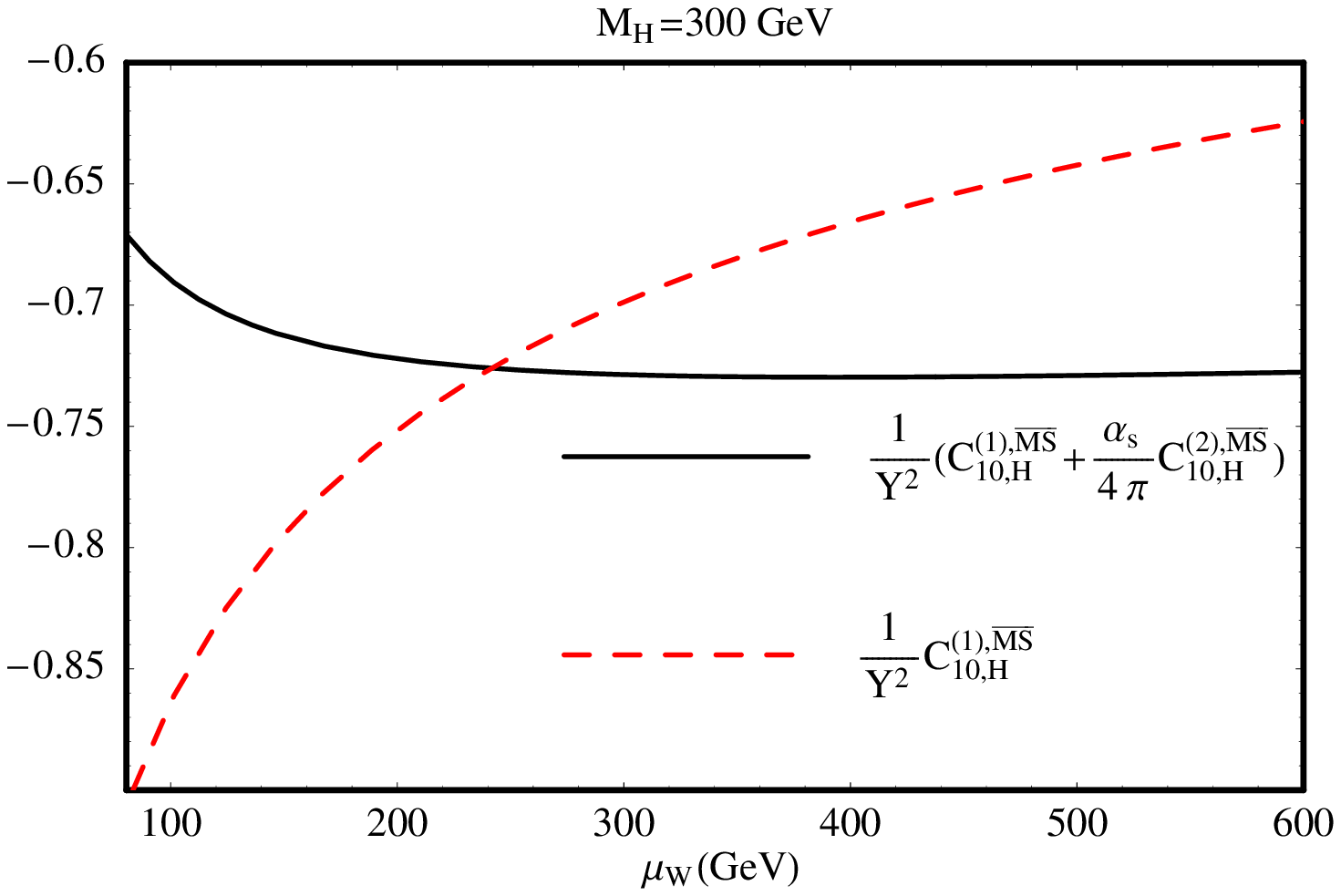,scale=1.0}
\end{center}
\caption{Dependence of the rescaled Wilson coefficient 
$\hat{C}_{10,\smallthdm}(\mu_b)$ on the matching scale $\mu_W$
(see eq. (\ref{rescaled})) for 
$M_H=300$ GeV.
The dashed line shows the one-loop contribution expressed in
\ms-scheme for the $t$-quark mass, 
while the solid line includes the two-loop corrections in the same
scheme.}
\label{combimuw}
\end{figure}

Looking at the renormalization group equation (RGE) \cite{Gambino3} 
for $\hat{C}_{10,\smallthdm}$, one finds that $\hat{C}_{10,\smallthdm}$
does not run, i.e. 
\begin{equation}
\hat{C}_{10,\smallthdm}(\mu_b)=\hat{C}_{10,\smallthdm}(\mu_W) \, ,
\end{equation}
where the low scale $\mu_b$ is of the order of $m_b$.  
In fig. \ref{combimuw} we show the dependence of 
$\hat{C}_{10,\smallthdm}(\mu_b)$ on the matching scale $\mu_W$ 
for $M_H=300$ GeV. It can be clearly seen that the inclusion 
of the two-loop contributions drastically lowers the dependence
on $\mu_W$. 
For $\mu_W >250$ GeV, $\hat{C}_{10,\smallthdm}(\mu_b)$ 
at two-loop precision is nearly $\mu_W$-independent. For $\mu_W$  
between $M_W$ and 250 GeV the two-loop Wilson coefficient varies  
about $\pm 4\%$, whereas the corresponding 
one-loop coefficient varies about $\pm 11 \%$. 

\vspace{1cm}

\noindent
To summarize: In this letter we have presented QCD corrections to 
the charged Higgs induced contributions to the Wilson coefficients 
$C_9(\mu_W)$ and $C_{10}(\mu_W)$ in type-I and type-II 2HDMs. 
These two-loop results are important ingredients
for complete NNLL calculations of various observables related to the
decay $B \to X_s l^+ l^-$ in these models.

\vspace{1cm}

\noindent
Just before submitting the present paper, we became aware of the
PhD thesis of Ch. Bobeth 
(http://tumb1.biblio.tu-muenchen.de/publ/diss/ph/2003/bobeth.pdf),
where the two-loop results for the charged Higgs boson contribution to
$C_9$ and $C_{10}$ are contained. We have
checked that our results agree.

\vspace{1cm}

\noindent
We would like to thank K. Bieri and D. Wyler for helpful discussions. S.S
would like to thank M. Misiak and J. Urban for fruitful discussions and 
advice regarding the technical details of the two-loop calculations. 
This work is partially supported by: the Swiss National Foundation; 
RTN, BBW-Contract No.~01.0357 and EC-Contract HPRN-CT-2002-00311 (EURIDICE).
\setlength {\baselineskip}{0.2in}
\bibliographystyle{unsrt}

\end{document}